# Intrusion Detection Systems for Flying Ad-hoc Networks

Jordan Quinn and Safdar Hussain Bouk
VMASC, Old Dominion University, 5115 Hampton Blvd, Norfolk, VA 23529.

*Abstract*— Unmanned Aerial Vehicles (UAVs) are becoming more dependent on mission success than ever. Due to their increase in demand, addressing security vulnerabilities to both UAVs and the Flying Ad-hoc Networks (FANET) they form is more important than ever. As the network traffic is communicated through open airwaves, this network of UAVs relies on monitoring applications known as Intrusion Detection Systems (IDS) to detect and mitigate attacks. This paper will survey current IDS systems that include machine learning techniques when combating various vulnerabilities and attacks from bad actors. This paper will be concluded with research challenges and future research directions in finding an effective IDS system that can handle cyber-attacks while meeting performance requirements.

Keywords— flying ad-hoc networks, unmanned aerial vehicles, machine learning, intrusion detection.

## I. INTRODUCTION

With the growing popularity of Unmanned Ariel Vehicles (UAVs), they have branched outside the military realm and into the hands of everyday civilians. Their usage has exploded into many industries worldwide and has completed diverse missions ranging from disaster management, search and rescue, weather monitoring, agricultural monitoring, and healthcare delivery [1-5].

With continuous technological improvements in UAVs, much progress has been made in creating wireless networks to accommodate their airborne missions. The latest form of such a network is the Flying Ad-hoc NETwork (FANET). FANET is a decentralized wireless network comprised of Unmanned Ariel Vehicles (UAVs), each representing nodes that communicate with each other while in flight [6],[7]. This also includes a Ground Control Station (GCS), which communicates with the other UAVs, starting with the closest drone in its proximity, as illustrated in Figure 1 [8]. While this wireless network protocol offers UAVs scalability, low latency, and resilience benefits, it is exposed to vulnerabilities other networks share [9]. These vulnerabilities make it necessary to explore automated solutions that effectively protect the FANET while not compromising the UAVs' performance.

This paper offers the following contributions below:
- A Survey of recent IDS using machine learning (ML) techniques on UAVs.
- Discuss research challenges and limitations.
- Suggested future research directions based on the research and lessons learned.

The remainder of this paper is arranged as follows: Section II defines FANET and UAV components. This will also mention security models. Section III addresses IDS and machine learning. Section IV is the survey of ML UAV IDS. Section V consists of research challenges. Section VI is the conclusion of this paper.

## II. FANET & UAV COMPONENTS

The FANET's primary purpose is to provide fast, dependable, and effective communication links between the UAVs without regard to any prior infrastructure placement. What makes FANETs attractive is the fact that they are inexpensive, fast to set up, scalable when adding numerous UAVs, and have high fault tolerance [10]. Although FANETs are commonly compared to the architecture of Mobile Ad-hoc NETworks (MANETs) and Vehicle Ad-hoc NETworks (VANETs), they have unique characteristics such as mobility, memory, and power when communicating with each other [11], [12], [13]. These characteristics affect not only the performance of the FANET but also the effectiveness of cyber security controls.

UAVs today can be categorized by numerous criteria such as purpose, shape, weight, and communication capabilities [14],[15]. Nevertheless, they all share similar components while operating in flight. For example, inside a popular multi-motored drone such as the DJI's Phantom 4, other drones can relate to having similar features such as types of sensors (GPS receivers, obstacle avoidance, etc.), motherboard, CPU, memory, battery power source, flight controller, and motors [16],[13]. With the given components come several challenges that must be addressed:

- **Power limitations**: Given UAVs components, battery power restricts the UAV capabilities. Also, Cyber-attacks can maliciously drain the power consumption of the UAV, causing it to disconnect from the FANET altogether [13].

- **Memory and CPU**: Robust cryptography methods are required within the UAVs to ensure data confidentiality. However, options are minimal due to limited battery power, hardware, and processing speeds. Without energy-efficient cryptographic algorithms, successful authentication attacks would exhaust UAV resources [13].

- **Mobility**: UAV movement constantly fluctuates, impacting the network topology and performance. For instance, UAV node speeds ranging from 30 to 460km per hour [18] make high mobility a significant challenge [19]. Due to these challenges, UAVs are subject to interference-type attacks such as signal jamming.

- **Wireless Network**: A communications medium must be established to transmit navigation and sensory data. FANETs may use wireless protocols such as IEEE 802.11, IEEE

802.15.4, Bluetooth, Satellite, or cellular mobile technologies, each with its challenges in signal strength, bandwidth, network management, and security [20].

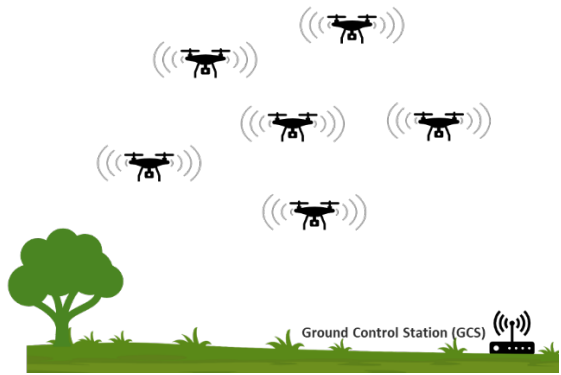

*Figure 1. FANET Structure Illustration*

FANET communication systems are also vulnerable to the same cyber-attacks as many wireless networks, such as jamming, spoofing, and intrusion of their network [32] [21]. However, the stakes have never been higher due to the mobility of drones and the safety concerns should they be compromised while in flight.

From a cybersecurity perspective, one must follow the most basic security model, such as the CIA triad, to search for vulnerabilities within any system. The CIA triad is an acronym for Confidentiality, Integrity, and Availability. Confidentiality ensures that data or information is viewable by authorized nodes. Integrity refers to communicating messages that are not altered or deleted by anyone or anything. Availability describes a node always being online and ready to receive instructions from other nodes [21]. With this security model in mind, security measures like an IDS system can be implemented to protect or detect malicious threats on the FANET network.

### III. IDS & MACHINE LEARNING

An IDS is a system composed of hardware or software that monitors network traffic looking for unauthorized behavior to report [22]. Types of unauthorized behavior include but are not limited to malware, UAV spoofing, routing attacks, and data forgery [23]. When deploying the IDS onto the network, it can be either host-based or network-based. For example, the host-based IDS system can be inserted directly inside each UAV, while the network-based IDS system can be inserted inside a network system on land. Given how decentralized FANETs are, and their component challenges, host-based IDS deployment may be the most effective implementation in detection. When it comes to the IDS detection methods, there are several types of techniques

1) Signature-based detection relies on comparing previous attack signatures or patterns to suspicious activity on the network to match the same signature. One major drawback is that it cannot detect newer attacks or zero-day exploits because there are no known signatures to compare it to.
2) Anomaly-based detection relies on a predefined model of normal network behavior instead of relying on patterns or signatures. Should network traffic not fit the model, it predicts the behavior as an anomaly. This method is excellent when detecting newer attacks but has limitations, such as seeing encrypted packets and higher false positives [22].
3) Hybrid detection is a combination of both signature and anomaly-based detection methods. This reduces the number of false positives while combining both their strengths [23].
4) Machine learning-based detection uses Machine learning (ML) algorithms to identify malicious network traffic. The detection methods can be signature, anomaly, or hybrid-specific. The main difference with ML is that the algorithms can reconfigure the IDS and improve the detection accuracy of newer or missed threats over time [24].

There have been attempts to create an IDS solution in FANET without machine learning, but they needed more overall effectiveness. For example, a proposed threat estimation model based on the belief approach was used to reduce IDS false positives. However, this method was based on known behaviors, limiting its effectiveness against unknown anomalies [25], [26]. Given the limited resources of UAVs and mobility demands on FANET, the survey findings in section IV will explore the machine learning-based IDS.

Machine learning is considered a subfield of artificial intelligence (AI), enabling computer systems to learn from experiences using algorithms and models over time. The algorithms are given sample data to build a model to make decisions or predictions outside the confines of its original programming [27].

Federated learning is a machine learning technique where the user's local data is never sent to centralized servers, ensuring data privacy [28]. Each client uses their data to train a piece of a model sent from a server on the ground. The client then encrypts its results and uploads data back to the ground server. The server then collects and decrypts all pieces of the model from all clients and pieces them together, forming an updated and improved global model to distribute back to the clients [29]. This process between the server and client is repeated until an acceptable accuracy is achieved before the final model is available for use by all clients.

While Federated Learning seems like the perfect IDS solution for FANET, it has never been applied to FANET specifically. However, Federated Learning has excellent promise as an effective IDS. Studies have shown that Federated learning effectively reports malicious traffic in other kinds of ad-hoc networks [22],[30].

### IV. SURVEY OF MACHINE LEARNING APPROACHES TO UAV SECURITY

Table 1 summarizes the UAV IDS approaches in building an automated system that can handle known and unknown attacks with the help of machine learning methods.

O. Bouhamed et al. [33] proposed IDS and intrusion detection prevention system (IDPS) using the method Deep

Reinforcement Learning (specifically Deep Q-learning) to allow autonomous detection of suspicious attacks on the UAV network, such as signal jamming and spoofing. Abu Al-Hhaija et al. [34] proposes an autonomous IDS that detects malicious threats against UAVs using deep convolutional neural networks (UAV-IDS-ConvNet). This was done using a two-class classifier from the UAV-IDS-2020 dataset to enhance detection from the deep-learning model.

Kyung Ho Park [35] proposed an IDS for UAVs leveraging unsupervised learning. He has pointed out that supervised learning models cannot identify attacks not included in machine learning models. Therefore, his model does require heavy data labeling but provides an effective IDS system for detecting jamming and spoofing attacks for UAVs. Liang Xiao[36] focuses on physical security methods to defend against jamming, eavesdropping, and spoofing attacks to UAV power supply. Reinforcement learning is proposed to achieve optimal power allocation against these attacks.

Gaoyang Liu et al. [37] was concerned with how bad actors easily imitate satellite signals and proposed a GPS detection system that uses machine learning algorithms Xgboost and K-NN to establish the actual position of the UAV and detect if its positional data has been compromised. Menaka Arthur et al. [38] proposes a lightweight IDS using an unsupervised featured algorithm, Self-Taught Learning (STL), which builds a training dataset from unlabeled data collected from the UAV's sensors. Multiclass SVM is also used to ensure high detection rates of the IDS.

Jason Whelan et al. [39] proposed one-class classifiers, such as One-class Support Vector Machine (OC-SVM) to train their "novelty-based" IDS, which only learns typical sensor values from previous flight logs. Roshaan Mehmood et al. [40] Proposed finding the most accurate machine learning algorithm from SVM, KNN, and Random Forest Classifier to train the model off the CIC-IDS2018 dataset. While using Cooja Simulator to simulate the UAV environment, Random Forest Classifier had the best accuracy rate ranging from 95%-96%. Rabie Ramadan et al. [41] proposed an IDS for FANET using deep learning and big data analytics. The framework used two Recurrent Neural Network (RNN) modules, one at ground level and another inside FANET.

## V. Research Challenges

With machine learning playing a more significant role in UAV IDS, effectiveness and efficiency seem obtainable. However, by adding ML into UAV IDS, some challenges must be addressed:

- Data collection and quality: training algorithms that detect network intrusions require mass volumes of data that are in short supply in this subject. Worse yet, the few available datasets may be incomplete or biased.
- Model selection: Many algorithms are available to develop a UAV IDS. However, selecting the "best" one depends on many factors, such as the selected data, desired levels of accuracy, and the problem that needs to be solved.
- Transferable: ML models are not universal to all UAV IDS. This is due to hardware and software requirements.
- Data security: Data can be manipulated or inserted into the IDS to become ineffective in identifying intrusions on UAV or the FANETs themselves. Ways to verify data integrity are a must in this case.

Addressing these challenges will require further research to ensure an efficient, effective, and secure UAV IDS system.

## VI. Conclusion

This paper provided a short overview of the usage of machine learning in UAV IDS. Also, discussions relating to FANET, UAVs, and ML were made. Lastly, research challenges and future research were addressed in creating an effective UAV IDS system.

| Reference | Proposed Scheme | Learning Method | Type of Attack |
| --- | --- | --- | --- |
| Omar Bouhamed[33] | Reinforcement learning on each UAV | Q-Learning: Framework ensures UAV integrity | Signal Jamming, Spoofing |
| Abu Al-Haija[34] | Supervised Learning. anomaly-based IDSs | Deep Learning: Detect anomalous Wi-Fi encrypted traffic | Signal Jamming |
| Kyung Park[35] | Unsupervised Learning, IDS leveraging autoencoder | Observed unlabeled attacks via simulations. | DoS Attack, GPS Spoofing |
| Liang Xiao[36] | Reinforcement learning, Physical-layer security | Q-learning, Prospect Theory to investigate attacks. | Jamming, Spoofing, Eavesdropping |
| Gaoyang Liu[37] | Supervised Learning, synchronization-free GPS IDS | K-Nearest Neighbor (K-NN), OpenSky Network for ATC data | GPS Spoofing |
| Menaka Arthur[38] | Semi-supervised learning, Self-healing IDS | Self-Taught Learning (STL), SVM is used to maintain accurate rates | GPS Spoofing, Jamming |
| Jason Whelan[39] | Supervised Learning, Novelty-based detection approach | SVM, learn from anomaly-free datasets | GPS Spoofing |
| Roshaan Mehmood[40] | Supervised Learning, | K-NN, SVM, Random Forest Classifier, | Jamming |
| Rabie Ramadan[41] | Supervised Learning, Distributed IDS framework | Deep Learning Based, RNN is used for detection. | Jamming |

TABLE I. SURVEY ON CURRENT MACHINE LEARNING IDS APPROACHES TO UAV SECURITY.